\documentclass[sigplan,screen]{acmart}
\usepackage{graphicx}
\usepackage{booktabs}
\usepackage{textcomp}
\usepackage{xcolor}
\usepackage{tikz}
\usepackage{enumitem}
\usepackage{multirow}
\usepackage{amsmath}
\usepackage{graphicx}
\usepackage{subfigure}
\usepackage{makecell}
\usepackage{amsmath}
\usepackage{mathtools}
\usepackage{amsthm}
\usepackage{hyperref}

\usepackage{soul}

\usepackage{xcolor}
\usepackage{colortbl}
\usepackage{booktabs}

\usepackage{algpseudocode}
\usepackage{algorithm}

\usepackage{setspace}

\newcommand*\circled[1]{\raisebox{.4pt}
                    {\tikz[baseline=(char.base)]{
            \node[shape=circle,draw,inner sep=1pt, style={fill=black, text=white}, scale=0.75] (char) {\textbf{#1}};}}}

\AtBeginDocument{%
  \providecommand\BibTeX{{%
    \normalfont B\kern-0.5em{\scshape i\kern-0.25em b}\kern-0.8em\TeX}}}

\setcopyright{acmcopyright}
\copyrightyear{2024} 
\acmYear{2024} 
\setcopyright{rightsretained} 
\acmConference[DAC '24]{61st ACM/IEEE Design Automation Conference}{June 23--27, 2024}{San Francisco, CA, USA}
\acmBooktitle{61st ACM/IEEE Design Automation Conference (DAC '24), June 23--27, 2024, San Francisco, CA, USA}
\acmDOI{10.1145/3649329.3656239}
\acmISBN{979-8-4007-0601-1/24/06}

\begin{document}

\title{  ICGMM: CXL-enabled Memory Expansion with \underline{I}ntelligent \underline{C}aching Using \underline{G}aussian \underline{M}ixture \underline{M}odel}

\author[1]{\normalsize Hanqiu Chen$^{1\S}$, Yitu Wang$^{2\S}$, Luis Vitorio Cargnini$^3$, Mohammadreza Soltaniyeh$^3$, Dongyang Li$^3$, Gongjin Sun$^3$, Pradeep Subedi$^3$, Andrew Chang$^3$, Yiran Chen$^2$ and Cong Hao$^1$
\\
{$^1$Georgia Institute of Technology, $^2$Duke University, and $^3$Samsung Semiconductor, Inc.}\\
{\normalsize \{hanqiu.chen, callie.hao\}@gatech.edu, \{yitu.wang, yiran.chen\}@duke.edu \\ 
\{v.cargnini, m.soltaniyeh, dongyang.li, gongjin.s, prad.subedi, andrew.c1\}@samsung.com}
}
 
\authornote{ $^{\S}$Equal contribution. Hanqiu Chen is responsible for prototype design on FPGA and paper writing, and Yitu Wang is responsible for GMM-based cache policy design. This work has been done as part of Hanqiu's and Yitu's internships in the Memory Solution Lab at Samsung Semiconductor, Inc. }


\renewcommand{\shortauthors}{Chen and Wang, et al.}


\begin{abstract}


 Compute Express Link (CXL) emerges as a solution for wide gap between computational speed and data communication rates among host and multiple devices. It fosters a unified and coherent memory space between host and CXL storage devices such as such as Solid-state drive (SSD) for memory expansion, with a corresponding DRAM implemented as the device cache. 
However, this introduces challenges such as substantial cache miss penalties, sub-optimal caching due to data access granularity mismatch between the  DRAM ``cache'' and SSD ``memory'', and inefficient hardware cache management.
To address these issues, we propose a novel solution, named ICGMM, which optimizes caching and eviction directly on hardware, employing a Gaussian Mixture Model (GMM)-based approach. We prototype our solution on an FPGA board, which demonstrates a noteworthy improvement compared to the classic Least Recently Used (LRU) cache strategy. We observe a decrease in the cache miss rate ranging from 0.32\% to 6.14\%, leading to a substantial 16.23\% to 39.14\% reduction in the average SSD access latency.
Furthermore, when compared to the state-of-the-art Long Short-Term Memory (LSTM)-based cache policies, our GMM algorithm on FPGA showcases an impressive latency reduction of over 10,000 times. Remarkably, this is achieved while demanding much fewer hardware resources.
\end{abstract}

\maketitle

\section{Introduction}
\label{sec:intro}

 Memory wall~\cite{gholami2021ai} is a critical performance bottleneck significantly impacts the hardware efficiency in memory-intensive tasks~\cite{kharya2021using,tang2022fedcor,wang2021rerec}.
Compute Express Link (CXL)~\cite{CXL2022} has emerged as a viable solution to this challenge. Built upon the serial PCI Express (PCIe) infrastructure, CXL offers a low-latency, high-bandwidth interconnect technology. It facilitates the direct sharing of memory and cache resources among devices, thus greatly enhancing the performance of data-intensive applications~\cite{sharma2022compute, li2023pond,wang2023ems,li2024ndrec}. 




Memory coherence provided by CXL facilitates memory expansion
to the storage of space of CXL devices, such as SSD.
Yang et al.~\cite{yang2023overcoming} proposed using DRAM as a ``cache'' for storage in CXL-enabled memory expansion systems, demonstrating improved SSD access efficiency, 
 however, DRAM ``cache'' still faces several challenges.~\circled{1} \textbf{Large cache miss penalties.} Compared with using DRAM as main memory and SRAM as cache, SSD data access latency is in microseconds~\cite{arpaci2018operating}, significantly higher than DRAM access in nanoseconds. This leads to substantial cache miss penalties.~\circled{2} \textbf{Suboptimal caching.} Suboptimal caching arises due to a mismatch in data access granularity between DRAM (64B) and SSD (4KB)~\cite{gal2005algorithms}. This necessitates a minimum cache block of one page size (4KB), often resulting in the caching of extraneous data.~\circled{3} \textbf{Hardware-inefficient cache policy designs}. Prior learning-based cache policies~\cite{zhong2018deep, shi2019applying,liu2020imitation,  yang2023gl} are managed by the software, inducing high overhead of executing the corresponding algorithm and overlooking the hardware performance considerations for CXL-device cache, i.e., the DRAM.

 
\begin{figure}[t!]
\centering
\subfigure{\includegraphics[width=1.00\linewidth]{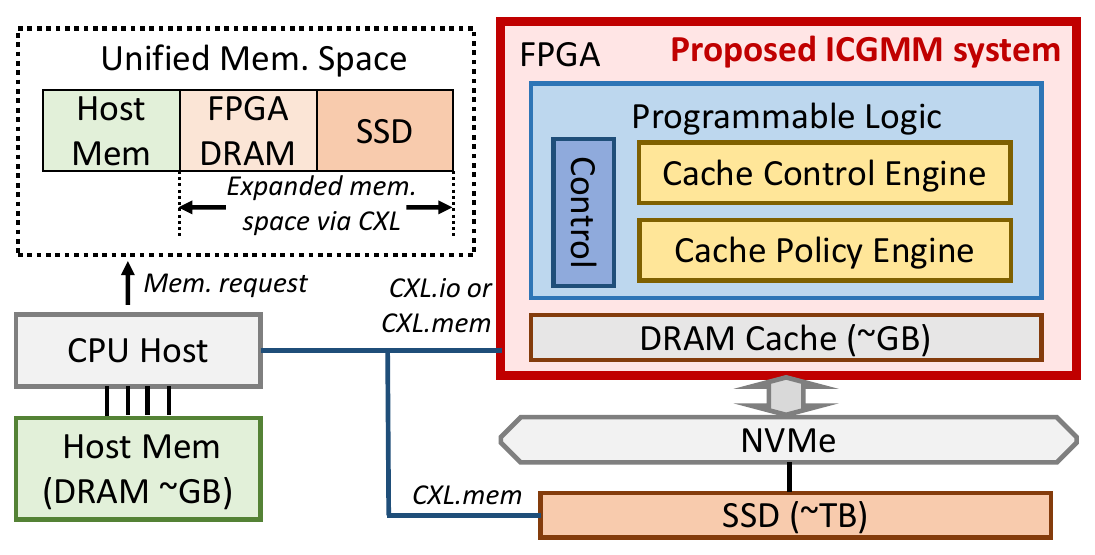}}
\vspace{-15pt}
\caption{CXL-enabled memory expansion. SSD serves as an extension of host main memory. FPGA DRAM is used as a cache to facilitate memory access to SSD via CXL. FPGA programmable logic is used for intelligent cache management. 
}
\vspace{-15pt}
\label{fig:HW_overview}
\end{figure}

Motivated by these challenges, in this paper, we propose \textbf{ICGMM, a hardware-managed DRAM caching system for CXL-enabled memory expansion} prototyped on FPGA. ICGMM incorporates a hardware-efficient cache policy engine based on the Gaussian Mixture Model (GMM) for \ul{\textit{improved cache hit rate and reduced average memory access latency}}. Our contributions can be summarized as follows in four aspects: system, algorithm, hardware, and evaluation.
\begin{enumerate}[leftmargin=*]
\item \textbf{System -- Hardware-managed DRAM cache system design.} ICGMM is an end-to-end DRAM cache management system prototyped on FPGA. It features a hardware-implemented cache controller that functions independently of the host control. ICGMM also incorporates the GMM algorithm implemented on hardware for 
intelligent caching and eviction.

\item \textbf{Algorithm -- GMM-based cache policy for intelligent caching and eviction.} 
We develop a two-dimensional GMM to predict the access frequency of requested SSD pages by leveraging physical addresses and timestamps from host memory access requests as the inputs. Based on GMM prediction, only frequently accessed pages will be cached in DRAM to reduce cache miss penalty, and the least frequently accessed pages will be evicted if needed.

\item  \textbf{Hardware -- Optimized GMM hardware design with dataflow architecture.}
Our GMM design is hardware-friendly because GMM computation is highly pipelineable, as each Gaussian function operates independently. Compared to LSTM-based cache policies, small GMM size and the capability of GMM to predict SSD page access frequency on-the-fly using current status trace information without tracing back previous information offers significant memory advantages. To further minimize GMM overhead and resolve the complex control, we also designed a dataflow architecture ensuring that GMM computation can be overlapped with SSD access.



\item \textbf{Evaluation -- On-board measurement.} 
We evaluate the performance of ICGMM end-to-end using the Alveo U50 FPGA. Compared to the traditional Least Recently Used (LRU) cache policy, ICGMM achieved a 0.32\% to 6.14\% decrease in cache miss rates, resulting in a 16.23\% to 39.14\% reduction in average SSD access latency across seven mainstream benchmarks encompassing various application types. Furthermore, GMM implementation outperforms LSTM-based policies on FPGA, reducing latency by over 10,000$\times$ while requiring fewer hardware resources.

\end{enumerate}

\section{Preliminary and Motivations}


\subsection{CXL-enabled memory expansion}
Compute Express Link (CXL)~\cite{CXL2022} is a novel interconnect protocol built on PCIe that connects heterogeneous devices within a unified memory space. Its capability of simply using host load/store instructions via CXL.io and CXL.mem protocol accelerates device memory access from the host. Using DRAM as a cache for storage devices like SSDs is a feasible memory expansion option for high-efficiency SSD access in CXL-enabled systems~\cite{yang2023overcoming}. However, as explained in Sec.~\ref{sec:intro}, the DRAM cache is inefficient in facilitating data transfer. It is necessary to align the DRAM cache block size to 4KB to be consistent with SSD minimum access granularity. This granularity mismatch can decrease cache hit rates due to the inclusion of infrequently accessed data.




Motivated by the \textbf{suboptimal coarse-grained caching and large cache miss penalties}, we propose to design a more powerful cache policy engine by predicting the frequently accessed data at the page level (4KB), to \textbf{improve the cache hit rate and thus largely reduce the average SSD access latency} from the host via CXL. Fig.~\ref{fig:HW_overview} illustrates an overview of CXL-enabled memory extension, where our proposed ICGMM prototyped on FPGA is highlighted.

\begin{figure}[t!]
\centering
\subfigure{\includegraphics[width=1\linewidth]{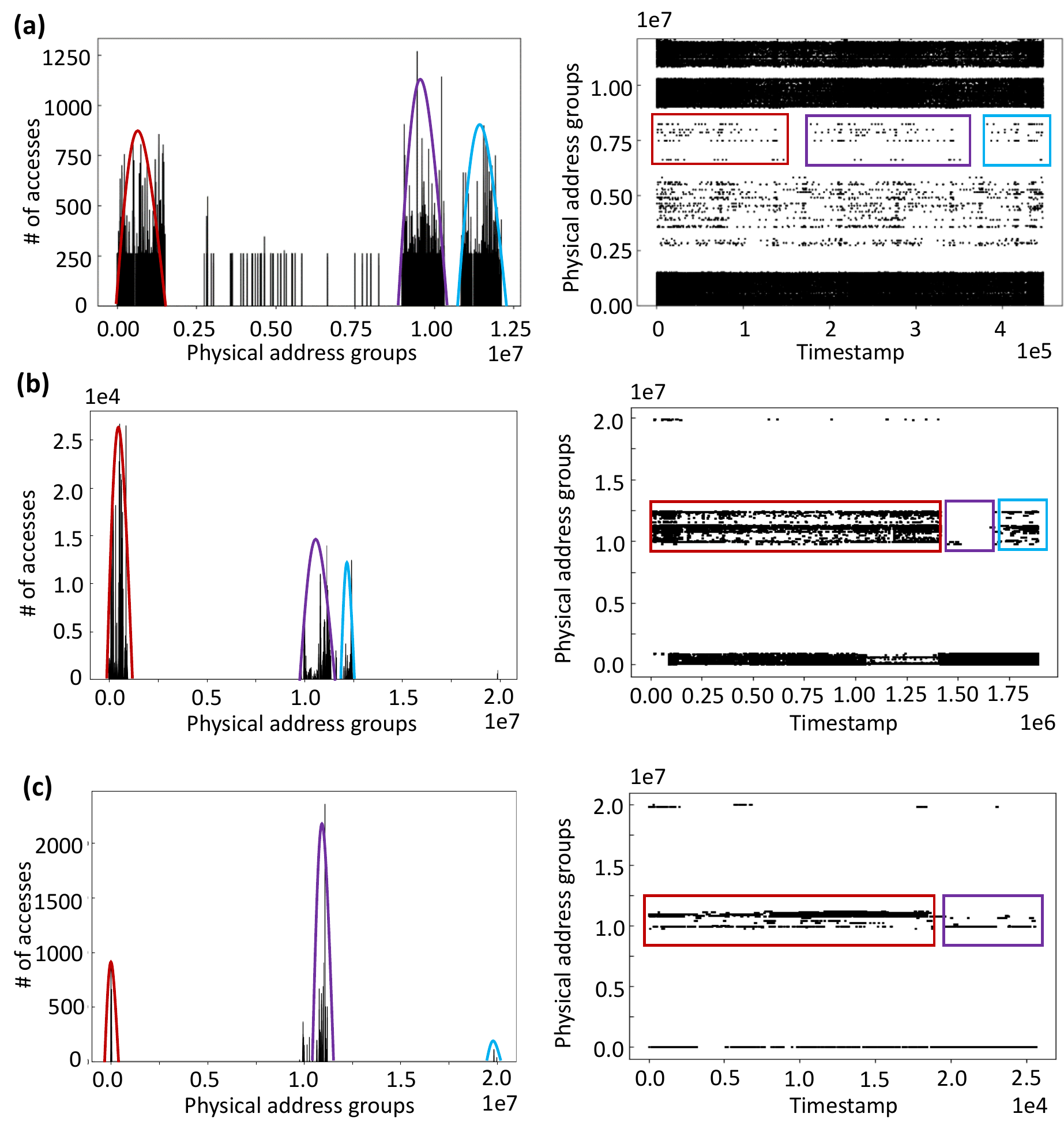}}
\vspace{-20pt}
\caption{Memory access spatial distribution (left) and temporal distribution (right) from three benchmarks: (a) dlrm~\cite{ArchImpl19}, (b) parsec~\cite{bienia2008parsec}, and (c) sysbench~\cite{sysbench}. Spatial distribution can be fitted with different Gaussian functions; temporal distribution shows uneven access frequency within a specific range of addresses (see colored annotations).}

\vspace{-15pt}
\label{fig:trace_analysis}
\end{figure}

\subsection{Learning-based cache policy}
Learning-based cache policies utilize historical caching data to forecast future caching priorities to boost cache hit rates. 
Integrating machine learning into cache policy design is currently a prominent research area. For example,
DeepCache~\cite{zhong2018deep}  and Glider~\cite{shi2019applying} innovate in smart caching with an LSTM-based cache line predictor.
LeCaR~\cite{vietri2018driving} and  PARROT~\cite{liu2020imitation} adopt imitation (IL) and reinforcement learning (RL) for more complex policies. GL-Cache~\cite{yang2023gl} groups similar cache objects for collective learning and eviction decisions.


While existing learning-based cache policies enhance cache hit rates, they are predominantly software-centric, neglecting the hardware cost for actual execution. In contrast, we propose a \textbf{hardware-friendly and lightweight} cache policy engine, aiming for \textbf{reduced hardware overhead and resource utilization} compared to existing cache policies.

\subsection{Gaussian Mixture Model (GMM)}
The GMM~\cite{reynolds2009gaussian} is a flexible clustering model for capturing the underlying structure of data through a combination of Gaussian distributions. We conduct a preliminary study on three trace benchmarks:  \textit{dlrm}~\cite{ArchImpl19}, parsec~\cite{bienia2008parsec} and sysbench~\cite{sysbench} with memory access spatial and temporal distribution patterns shown in Fig.~\ref{fig:trace_analysis}: the \textit{spatial distribution} describes the memory access frequency associated with data location, while the \textit{temporal distribution} describes memory access location associated with time. We find that \textit{spatial distribution} aligns with a mixture of Gaussian distributions, each having distinct means and co-variances. \textit{Temporal distribution} is non-random and can be clustered into multiple groups, which is suitable to use GMM for clustering. We also observe similar spatial and temporal patterns on other trace benchmarks. To obtain the optimal accuracy, GMM needs to combine these two distributions for prediction because although some specific address ranges have higher access frequency than others, however, the access frequency distribution is \textit{uneven} in temporal. Only considering spatial distribution will degrade GMM prediction performance.

\begin{figure}[t!]
\centering
\subfigure{\includegraphics[width=1\linewidth]{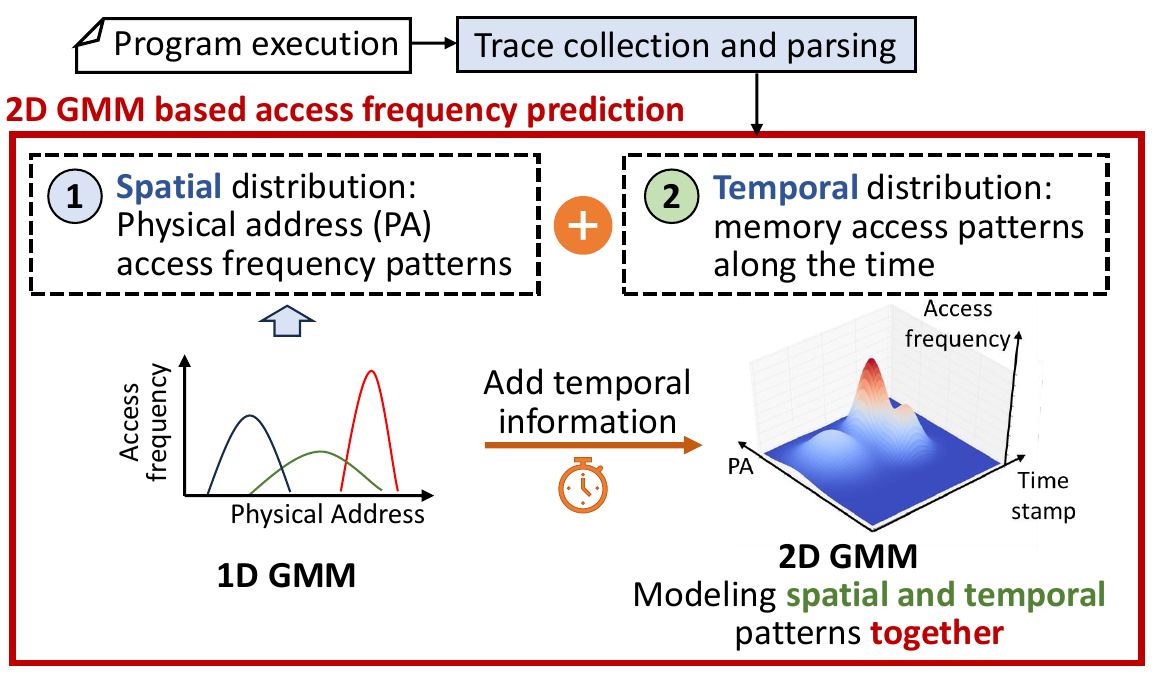}}
\vspace{-20pt}
\caption{We propose a two-dimensional GMM to capture both spatial and temporal memory access patterns. }

\vspace{-10pt}
\label{fig:SW_overview}
\end{figure}

Motivated by the \textbf{natural fit for employing GMM in modeling both spatial and temporal memory access patterns together}, we propose to use a \textbf{two-dimensional GMM} for memory access modeling, using transformed physical address ($P$) and timestamp ($T$) as inputs, as shown in Fig.~\ref{fig:SW_overview}. When extending to 2D, the Gaussian distribution can be expressed using the following equations:
\begin{equation}
\small
\mathcal{N}(\mathbf{x},|\boldsymbol{\mu}_{k}, \boldsymbol{\Sigma}_k) = \frac{1}{2\pi|\boldsymbol{\Sigma}_k|^{\frac{1}{2}}} \exp\left(-\frac{1}{2}(\mathbf{x}-\boldsymbol{\mu}_k)^T\boldsymbol{\Sigma}_k^{-1}(\mathbf{x}-\boldsymbol{\mu}_k)\right)
\end{equation}
\begin{equation}
\mathbf{x} = \begin{pmatrix} P \\ T \end{pmatrix} \quad \boldsymbol{\mu}_k = \begin{pmatrix} \mu_P \\ \mu_T \end{pmatrix}_k \quad \boldsymbol{\Sigma}_k = \begin{pmatrix} \sigma_{PP} & \sigma_{PT} \\ \sigma_{TP} & \sigma_{TT} \end{pmatrix}_k
\end{equation}
where $\boldsymbol{\mu}_k$ is a 2D mean vector and $\boldsymbol{\Sigma}_k$ is a $2\times2$ co-variance matrix of each 2D Gaussian function. By mixing $K$ Gaussian functions together using different normalized weights $\pi_k$ ($0 \leq \pi_k \leq 1, \sum_{k=1}^{K} \pi_k = 1$), our 2D GMM outputs a score $\mathcal{G}$ that predicts the future access frequency of each physical address, as shown in the following equation:
\begin{equation}
\small
\mathcal{G}(\boldsymbol{\pi}, \boldsymbol{\mu}, \boldsymbol{\Sigma}) = \sum_{k=1}^{K} \pi_k\mathcal{N}(\mathbf{x}|\boldsymbol{\mu}_k, \boldsymbol{\Sigma}_k)
\end{equation}
where $\boldsymbol{\pi}$, $\boldsymbol{\mu}$, and $\boldsymbol{\Sigma}$ are trainable parameters for encoding different types of memory access traces.

\begin{algorithm}[t]
\small
\caption{Trace timestamp transformation for GMM}

\label{Algorithm: trace preprocessing}

\begin{algorithmic}[1] 
\State $timestamp \gets 0$
\State $index \gets 0$
\While{a memory request comes}
    \If{$index \geq \text{len\_window}$}
        \State $timestamp \gets timestamp + 1$
        \State $index \gets 0$
    \EndIf
    \If{$timestamp \geq \text{len\_access\_shot}$}
        \State $timestamp \gets 0$
    \EndIf
    \State $index \gets index + 1$
\EndWhile
\end{algorithmic}

\end{algorithm}

\section{Cache Policy Engine Design with GMM}

\begin{figure}[t!]
\centering
\subfigure{\includegraphics[width=1\linewidth]{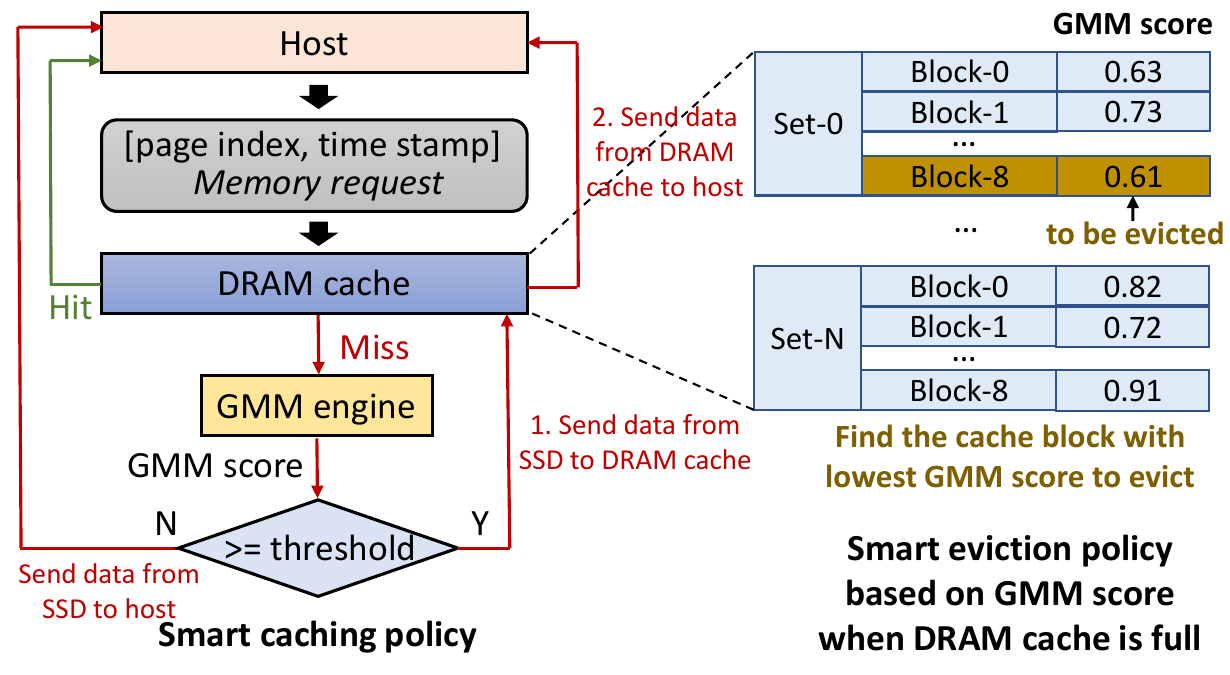}}
\vspace{-20pt}
\caption{Intelligent caching and eviction with GMM.
}
\vspace{-15pt}
\label{fig:Intelligent caching and eviction with GMM}
\end{figure}

The cache policy engine, integrated with GMM, determines whether the current requested SSD page will be cached and which page will be evicted based on GMM prediction score for access frequency. A higher score indicates a higher likelihood of future page access.  As shown in Fig.~\ref{fig:Intelligent caching and eviction with GMM}, when the score is higher than the threshold, the page will be cached in the DRAM. To gather sufficient data for GMM training, each program runs for a long time, enough until passing the warm-up stage and the memory access pattern is stable. 
We use an open-sourced tool~\cite{yang2023overcoming} for trace collection,  including the read/write information, physical address, and access time. Then the trace will be parsed and processed (Sec.~\ref{Sec.trace processing}) to break down into meaningful time windows and exact inputs for GMM. The input of GMM includes the page index calculated from the physical address and transformed access timestamp. Next, GMM will make caching and eviction decisions based on the GMM score (Sec.~\ref{Sec.Intelligent caching and eviction with GMM}). GMM is trained using the Expectation-Maximization (EM) algorithm (Sec.~\ref{sec:GMM training}).

 


\subsection{Trace processing}
\label{Sec.trace processing}

We do a preprocessing for the trace to facilitate GMM training. To mitigate program warmup biases, we discard the initial 20\% and final 10\% of traces. Since SSD access granularity is in 4KB pages, differing from host access granularity, we consolidate addresses into pages by assigning a page index (PI) computed from the physical address (PA) as $PI = PA << 12$. 

To help GMM capture memory access locality better,
we first partition the whole trace into small segments named \textit{access shots}. The number of traces inside one access shot is represented by $len\_access\_shot$. We then continue to divide the traces inside one access shot into much smaller segments named \textit{time window}. The number of traces inside a time window is represented by $len\_window$. As outlined in Algorithm~\ref{Algorithm: trace preprocessing}, we assign the same timestamp to traces inside the same time window, and the time window can be indexed by this timestamp. Different time windows are indexed with an incremental timestamp, which will be reset to zero if a trace reaches the end of the access shot. In our experiments, we empirically choose $len\_window = 32$ and $len\_access\_shot = 10,000$ for optimal GMM training performance.



\begin{figure*}[ht!]
\centering
\vspace{-12pt}
\subfigure{\includegraphics[width=1.00\linewidth]{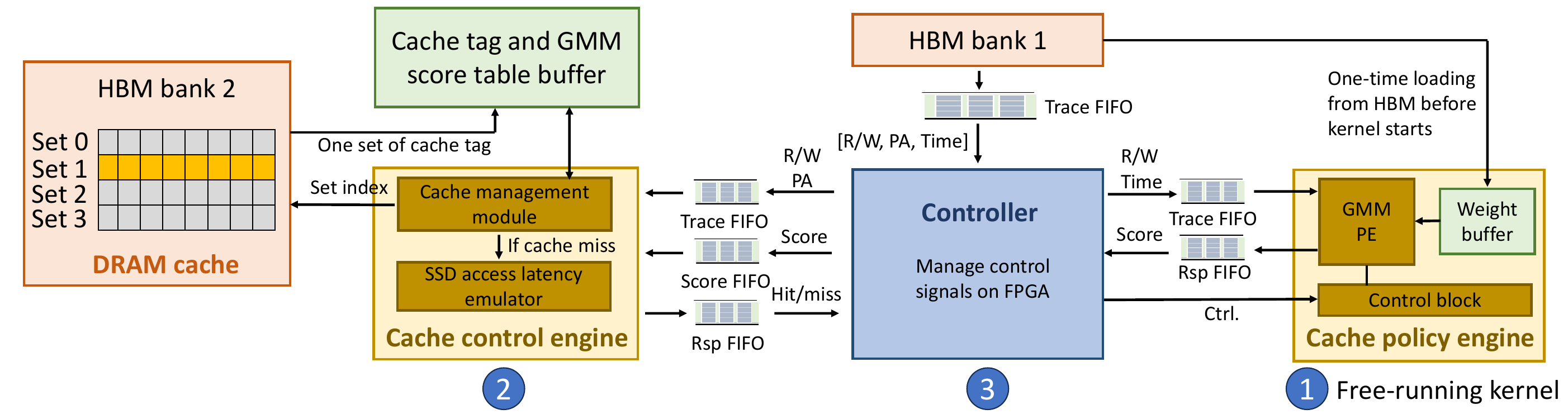}}
\vspace{-25pt}
\caption{ICGMM hardware architecture design with three main modules: cache control engine, cache policy engine, and signal controller. ICGMM is designed as a dataflow architecture with FIFO interfaces between different modules for high parallelism and efficient data-driven control. PA means the physical address. 
\vspace{-10pt}
}
\label{fig:HW implementation}
\end{figure*}

\subsection{Intelligent caching and eviction with GMM}
\label{Sec.Intelligent caching and eviction with GMM}
As shown in Fig.~\ref{fig:Intelligent caching and eviction with GMM}, to utilize GMM for intelligent caching, when an application is running on our system, if the page index of a memory request hits in the DRAM cache, the data is sent directly to the host, bypassing the GMM. On a cache miss, the GMM calculates the score for that page; if the score falls below a certain threshold, which indicates infrequent future access, then the page won't be cached and is sent directly from the SSD to the host. In contrast, scores above the threshold trigger caching those pages to the DRAM, anticipating frequent access in the future.

We also utilize GMM for intelligent eviction by substituting the LRU counter with the GMM score within cache blocks. Upon a full DRAM cache necessitating eviction, we sort the blocks by GMM score within the relevant set and evict the block with the lowest score, indicative of infrequent future access. After eviction, we update the cache with a new page from SSD, along with its corresponding GMM score.

\subsection{GMM training}
\label{sec:GMM training}
GMM training is conducted through unsupervised learning, utilizing the Expectation-Maximization (EM) algorithm.
 This process comprises two main steps. In the first expectation step, the probability of each trace belonging to each Gaussian function is calculated based on Bayes' theorem.
In the second maximization step, GMM parameters ($\boldsymbol{\pi}$, $\boldsymbol{\mu}$ and  $\boldsymbol{\Sigma}$) are updated, ensuring that GMM parameters better represent input trace memory access patterns. After each iteration, the convergence is checked by calculating the change in maximum likelihood estimate (MLE) of $\boldsymbol{\pi}$, $\boldsymbol{\mu}$ and  $\boldsymbol{\Sigma}$. If the change is below the predefined threshold, GMM is converged and parameters will be saved for inference.

\section{ICGMM Hardware Architecture Design}

We prototype ICGMM on FPGA not only to demonstrate its hardware friendliness but, more importantly, ICGMM opens a new opportunity of utilizing the SmartSSD~\cite{lee2020smartssd} as a CXL-enabled memory expansion device: SmartSSD has a small FPGA attached to its storage, which can function as a cache and controller for the SSD. Since host-based DRAM cache management is inefficient due to the high latency from frequent data exchanges between the host and SmartSSD, a hardware-managed cache control through programmable logic in the SmartSSD can significantly improve the efficiency of using DRAM as a cache for SSD.

However, the hardware implementation of ICGMM is non-trivial because of the complex control required over various modules and blocks. For example, the cache policy engine must be constantly active to wait for signals, triggering computations only upon cache miss. This run-time dynamic control cannot be determined during the hardware design phase. Therefore, we propose dataflow architecture using ``free running kernels'' for data-driven control and high parallelism between different modules and blocks.




ICGMM is prototyped on Xilinx Alveo U50 FPGA, utilizing its high-bandwidth memory (HBM) as the DRAM cache. As shown in Fig.~\ref{fig:HW implementation}, the system comprises three main modules: ~\circled{1} \textbf{a cache policy engine} with an optimized GMM kernel for predicting the likelihood of future page access frequency; ~\circled{2} \textbf{a cache control engine} responsible for cache management, hit/miss determination, and cache replacement, which also includes a SSD access latency emulator;  ~\circled{3} \textbf{a signal controller} for interfacing with cache control and policy engines 
and managing data flow between HBM, on-board buffers, and different modules and blocks.

\subsection{GMM policy engine design}
Our GMM-based cache policy engine is an independent, data-driven module that can continuously process data without external intervention, which is also known as a ``free-running kernel''. Upon detecting a new trace from trace FIFO, the engine retrieves and decodes the trace to GMM inputs format and executes GMM inference. 
A dedicated control block 
oversees this policy engine, awaiting directives from a central signal controller. If the signal controller activates the cache policy engine, the trace FIFO and response (rsp) FIFO are activated, and GMM scores will be returned from the rsp FIFO to intelligently manage the cache. Otherwise, two FIFOs will be closed, and the system will run a default traditional Least Recently Used (LRU) strategy for cache replacement. 

For optimized hardware performance, we exploit the feature of independent Gaussian functions in GMM by constructing a deep computation pipeline that processes different stages of the computation simultaneously with an initiation interval (II) = 1. For the task of score accumulation from different Gaussian functions, which requires sequential processing, we implement a shift register that holds temporal values during accumulation to resolve the data dependency. Furthermore, the GMM size is small enough to be stored within an on-board weight buffer, which avoids continuous data exchanges between the HBM and weight buffer.


\subsection{Cache control engine design}
The cache control engine consists of two modules, one for cache management, and another for emulating the SSD access latency. Within the cache management module, incoming traces from the trace FIFO queue are decoded to extract the set index, which is used to identify the appropriate cache set within the HBM. Instead of transferring the actual data in the cache, only the cache tags and GMM scores are transferred to the on-board buffer. Partitioning the cache tag and GMM score table buffer allows for simultaneous comparison of all tags from various blocks with the target tag of the incoming memory request, as opposed to the traditional sequential comparison. This parallel processing significantly accelerates cache hit/miss determination.


For more precise end-to-end ICGMM performance evaluation on FPGA, we also incorporate an SSD access latency emulator within the cache control engine. This emulator operates during a cache miss event, pausing the dataflow in the cache control engine for a set duration to emulate SSD response times. 
The parameters for response time vary according to the type of SSD or other storage devices.

\subsection{Dataflow architecture}
We employ a dataflow architecture to minimize the additional overhead brought by the cache policy engine and trace loading from HBM. 
 When the cache control engine is conducting tag comparison, new traces can be fetched from HBM into on-board registers, allowing simultaneous trace loading and cache management. Moreover, during a cache miss, the cache policy engine and SSD access emulator are triggered concurrently, significantly reducing extra delays caused by the cache policy engine. Moreover, dataflow architecture simplifies the complex control as data-driven control for "free-running" cache policy engine, resolving the non-deterministic behavior during run-time.
 


\section{Experiment}
\subsection{Experiment setup}

\textbf{Trace collection source.} For a comprehensive evaluation of ICGMM, we use different benchmarks, including both synthetic traces and traces from real-world applications. The synthetic trace benchmarks we choose are \textit{hashmap} and \textit{heap}~\cite{yang2023overcoming}. The real-world trace benchmarks are from different domains, including \textit{dlrm}~\cite{ArchImpl19} from deep learning recommendation systems, \textit{parsec}~\cite{bienia2008parsec} and \textit{stream}~\cite{mccalpin2006stream} from high-performance computing, \textit{memtier}~\cite{memtier_benchmark} and \textit{sysbench}~\cite{sysbench} from database systems.

\noindent
\textbf{Hardware deployment.} As a case study, the cache configuration we chose is shown as follows: cache size = 64 MB, block size = 4 KB, and associativity = 8. The number of Gaussian functions for GMM is 256. The target SSD for access latency emulation is fabricated using TLC technology, with average read latency = $75\mu s$ and write latency = $900\mu s$~\cite{arpaci2018operating}. ICGMM is deployed on Xilinx Alevo U50 FPGA running at $233 MHz$, using High-Level Synthesis (HLS) by Vitis HLS and Vivado 2023.1, with only 190 (14\%) BRAM and 117 (2\%) DSP consumption. 



\begin{figure}[t!]
\centering
\subfigure{\includegraphics[width=1\linewidth]{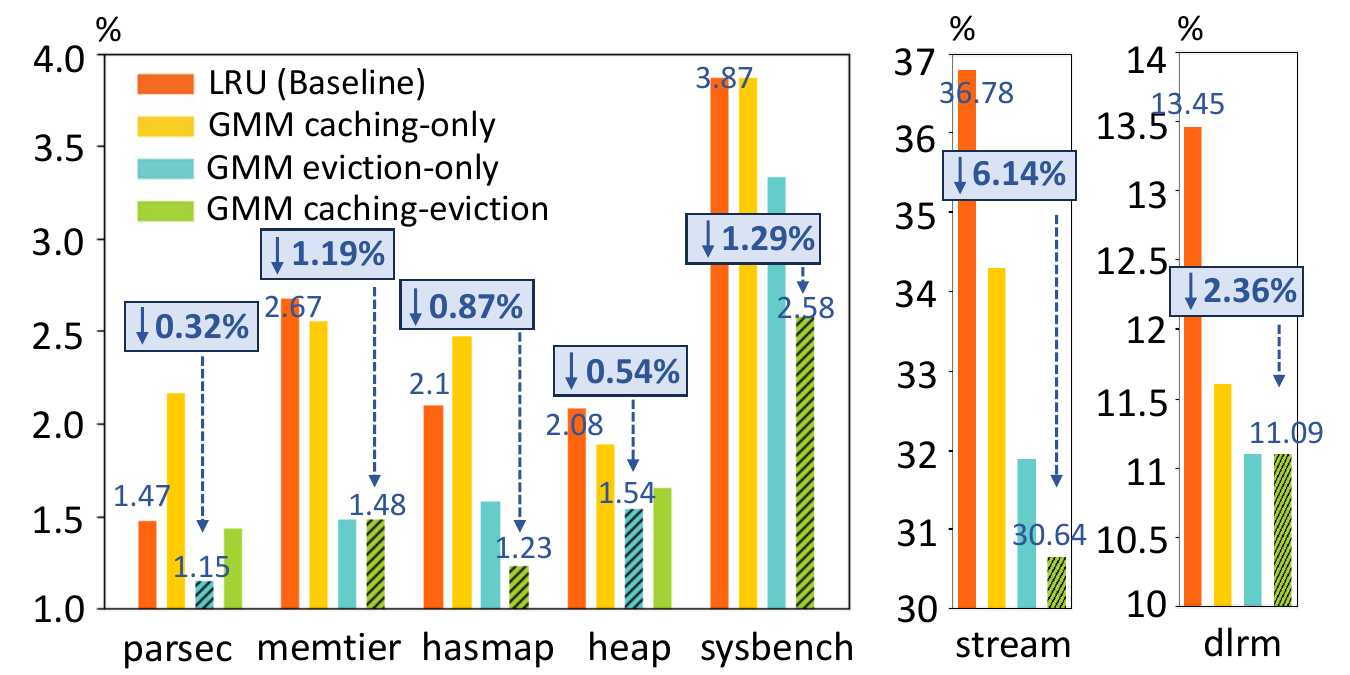}}
\vspace{-15pt}
\caption{Comparison of cache miss rates between baseline LRU and three distinct strategies using GMM as the policy engine.  Values the lower the better. Dashed bars indicate the best policy results.
}
\vspace{-15pt}
\label{fig:Cache miss rate}
\end{figure}

\subsection{Evaluation on GMM for caching and eviction}

We evaluate the effectiveness of GMM as a cache policy engine by comparing the cache miss rate of GMM against the baseline LRU policy with three strategies: GMM for smart caching, GMM for smart eviction, and a combination of both. We pick the strategy yielding the lowest cache miss rate, as shown in Fig.~\ref{fig:Cache miss rate}. Our results indicate that GMM reduces cache misses across all traces. Specifically, GMM eviction alone is the best for \textit{parsec} and \textit{heap}, while a combined approach achieves optimal performance for other traces.


\subsection{Hardware performance evaluation on FPGA }

\textbf{Average memory access latency reduction.}
Utilizing GMM for cache policy significantly lowers average memory access time by mitigating the substantial latency penalties of SSD access when cache miss happens. We conduct end-to-end on-board measurements, which reveal that DRAM cache hit time is $1 \mu s$. Upon a cache miss, GMM inference latency is $3\mu s$, which is quick enough to be overlapped with the SSD read ($75\mu s$) or write ($900\mu s$) request latency. In this case, cache miss penalties come from SSD access latency, with a total penalty reaching $75\mu s$ if SSD read is required and $975\mu s$ for dirty cache block writing back to SSD upon eviction. As shown in Table.~\ref{tab:benchmark-results}, GMM achieves a $16.23\%$ to $39.14\%$ reduction in average memory access time across seven benchmarks, compared to LRU.

\begin{table}[ht]
  \centering
  \small
  
  \caption{Comparison of average SSD access time using different cache policies: LRU and GMM.}
  \vspace{-5pt}
  \begin{tabular}{cccc}
    \toprule
    Benchmark & LRU & \textbf{GMM} & \textbf{Reduction (\%)} \\
    \midrule
    parsec & 3.92$\mu s$ & \textbf{3.29$\boldsymbol{\mu s}$} & \textbf{16.23} \\
    memtier & 2.98$\mu s$ & \textbf{2.09$\boldsymbol{\mu s}$} & \textbf{29.87} \\
    hashmap & 18.10$\mu s$ & \textbf{11.02$\boldsymbol{\mu s}$} & \textbf{39.14} \\
    heap & 16.48$\mu s$ & \textbf{12.46$\boldsymbol{\mu s}$} & \textbf{24.39} \\
    sysbench & 3.87$\mu s$ & \textbf{2.91$\boldsymbol{\mu s}$} & \textbf{24.79} \\
    stream & 156.39$\mu s$ & \textbf{125.71$\boldsymbol{\mu s}$} & \textbf{19.62} \\
    dlrm & 70.65$\mu s$ & \textbf{58.43$\boldsymbol{\mu s}$} & \textbf{17.30} \\
    \bottomrule
  \end{tabular}
  \label{tab:benchmark-results}
\end{table}

\begin{table}[ht]
\centering
\small
\caption{Resource utilization and latency comparison between LSTM and GMM for cache policy engine.}
\vspace{-5pt}
\begin{tabular}{cccccc}
\toprule
 & BRAM & DSP & LUT & FF & Latency \\
\midrule
LSTM & 339 & 145 & 85029 & 103561 & 46.3ms \\
\textbf{GMM} & \textbf{8} & \textbf{113} & \textbf{58353} & \textbf{152583} & \textbf{3$\boldsymbol{\mu}$s}
 \\
GMM gain & \cellcolor{yellow!50}2\% & 78\% & 69\% & 147\% & \cellcolor{yellow!50}15433$\times$ \\
\bottomrule
\end{tabular}
\label{tab:LSTM and GMM comparison}
\end{table}

\noindent
\textbf{GMM latency and resource utilization benefits.}
To highlight the hardware efficiency of GMM as a policy engine, we contrast it with an LSTM-based policy engine, which is commonly used in previous research~\cite{zhong2018deep, shi2019applying}. We design a three-layer LSTM model as a baseline with hidden dimension = 128, input sequence length = 32, and deploy the LSTM on the same FPGA platform with reasonable optimizations and similar DSPs utilization to ensure comparison fairness. During training, the LSTM is hard to converge across the same traces used for GMM using the same inputs, because it is unable to encode extensive temporal information in long traces with a lightweight design. A heavier LSTM model has a better encoding quality but at the cost of hardware efficiency. In comparison, as shown in Table~\ref{tab:LSTM and GMM comparison}, our LSTM baseline design is over 10,000$\times$ slower and consumes over 40$\times$  more BRAM than GMM, underscoring the superior hardware efficiency of GMM.

\section{Conclusion}

In this paper, we propose \textbf{ICGMM}, an intelligent hardware caching solution using a Gaussian Mixture Model (GMM) as a policy engine for CXL-enabled memory expansion systems. The optimized GMM-based policy engine smartly manages caching and eviction while minimizing overhead through dataflow architecture. ICGMM reduces cache miss rates by 0.32\% to 6.14\% and average SSD access time by 16.23\% to 39.14\% across diverse benchmarks. Furthermore, compared to the commonly used LSTM-based policy engine, the GMM-based policy engine achieves only 2\% on-chip memory usage and over $10,000 \times$ quicker inference speed.

\section{Acknowledgements}
This work is partly supported by the Samsung Memory Solution Lab University Program. Andrew Chang provided supervision. Luis Vitorio Cargnini guided the design of a GMM-based cache policy engine, and Mohammadreza Soltaniyeh guided the implementation of a hardware prototype. Dongyang Li, Gongjin Sun, Pradeep
Subedi, Yiran Chen, and Cong Hao contributed to insightful discussions and assisted in paper writing.



\bibliography{gatech}

\end{document}